\documentclass[a4paper]{article}
\usepackage{hyperref}
\usepackage{a4wide}
\usepackage{amssymb}
\usepackage{amsmath}
\usepackage{amsthm}
\usepackage{graphicx}
\usepackage{fancyhdr}
\usepackage{datetime}
\ddmmyyyydate
\begin{document}

\title{Generalized TBA and generalized Gibbs}
\author{Jorn Mossel and Jean-S\'ebastien Caux \\
{\it \small Institute for Theoretical Physics, Universiteit van Amsterdam,
Science Park 904,} \\ 
{\it \small Postbus 94485, 1090 GL Amsterdam, The Netherlands}}
\date{\today}

\maketitle
\begin{abstract}
We consider the extension of the thermodynamic Bethe Ansatz (TBA) to cases in which additional terms involving
higher conserved charges are added to the Hamiltonian, or in which a distinction is made between the
Hamiltonian used for time evolution and that used for defining the density matrix. Writing down equations describing the 
saddle-point (pseudo-equilibrium) state of the infinite system, we prove the existence and uniqueness of
solutions provided simple requirements are met. We show how a knowledge of the saddle-point rapidity
distribution is equivalent to that of all generalized inverse temperatures, and how the standard 
equilibrium equations for {\it e.g.} excitations are simply generalized. 
\end{abstract}

\section{Introduction}
Exactly solvable models of quantum mechanics are such that the basic Hamiltonian $H_0$ 
enforcing time evolution of the theory is part of an extended family of conserved charges $\hat{Q}_n$ in involution,
\begin{equation}
\left[ H_0, \hat{Q}_n\right] = 0, \hspace{5mm} \left[ \hat{Q}_n, \hat{Q}_m \right] = 0, ~~\forall ~n, m.
\end{equation}
Physically interesting cases correspond to charges expressed as integrals of (quasi-)local operators
and the existence of this set ultimately enforces the notion of factorized scattering associated with integrability.
Simple charges usually include a number operator $\hat{N} \equiv \hat{Q}_0$, the
total momentum operator $\hat{P} \equiv \hat{Q}_1$ and obviously the Hamiltonian itself, $H_0 \equiv \hat{Q}_2$. The rest,
$\hat{Q}_n$, $n > 2$ are not usually explicitly invoked or used
(although they are in principle available from the Algebraic Bethe Ansatz \cite{Korepin1993} via trace identities),
partly due to the fact that expressions for higher conserved charges quickly become unwieldy as one moves up their hierarchy.

It is a completely trivial fact that any functional combination of conserved
charges is itself a conserved charge. Viewing the charges as the `basis vectors' of the space of 
conserved charges, one can thus consider the generic expression
\begin{equation}
H(\{ \beta \}) = \sum_{n} \beta_n \hat{Q}_n
\label{eq:Hgeneric}
\end{equation}
as defining a generalized Hamiltonian depending on a (potentially infinite) set of real parameters $\beta_n$.
The interest of such constructions is twofold. First, and most obviously, it permits to easily extend
the `phase space' of integrable Hamiltonians considered, by using the charges as representatives of
competing interactions \cite{Tsvelik_1990,Frahm_1992,Muramoto_1999,Zvyagin_2001, Zvyagin_2003, Trippe_2007}, most typically by including only
one or a few additional charges as compared to the basic Hamiltonian $H_0$. 
Second, there exist situations in which one would like to consider functional averages (expectation values)
using a density matrix which is not given by the usual Gibbs weight involving the Schr\"odinger Hamiltonian
$H_S$ used for time evolution, but rather with a more general density matrix defined as \cite{Jaynes_1957,Rigol_PRL_2007}
\begin{equation}
\hat{\rho}_{GGE} = e^{-\sum_{n} \beta_n \hat{Q}_n }
\end{equation}
which is now generally known as the generalized Gibbs ensemble (GGE), and which purports to represent the
large-time density matrix resulting from a quench into an integrable system. In this case the generalized
(inverse) temperatures $\beta_n$ are set by somehow solving the self-consistency problem for the initial conditions 
\begin{equation}
\langle \hat{Q}_m \rangle = \text{Tr}\left\{ \hat{Q}_m e^{-\sum_{n} \beta_n \hat{Q}_n } \right\} /\mathcal{Z}_{GGE} \quad m=0,1,2,\ldots
\label{eq:GGEinit}
\end{equation}
in which $\mathcal{Z}_{GGE} = \text{Tr} e^{-\sum_{n} \beta_n \hat{Q}_n } $ is the generalized partition function.
Precisely how this is done depends on the situation; for free models it can be done explicitly by exploiting
the factorization into separate momentum sectors (see for example \cite{Rigol_PRL_2007,Calabrese_2007_JSTAT,2008_Barthel_PRL_100,2010_Fioretto_NJP_12,2011_Calabrese_PRL_106}). In the generic interacting case the  trace in \eqref{eq:GGEinit} cannot be taken explicitly. However, in the thermodynamic limit it is justified to consider the saddle-point of the partition function, which we will derive in the next section for the Lieb-Liniger model. On the other hand, defining the charges themselves and performing the averages in (\ref{eq:GGEinit}) typically represent an insurmountable problem.

It is our purpose here to provide a rather simple extension of the well-known thermodynamic Bethe
Ansatz (TBA) to these situations, and to discuss the existence of solutions and the actual implementation of calculations
in these cases.

\section{Generalized TBA for the Lieb-Liniger model}
For concreteness, we will illustrate the ideas by considering the Lieb-Liniger model, whose Hamiltonian is
\begin{equation}
H_0 = \int_{0}^{L} dx \left\{ \partial_x \Psi^\dagger(x) \partial_x \Psi(x) + c \Psi^\dagger(x) \Psi^\dagger(x) \Psi(x) \Psi(x) \right\}.
\label{eq:HLiebLin}
\end{equation}
We restrict ourselves to the case of positive coupling constant $c>0$, in which energies per unit length remain
finite in the thermodynamic limit. As is well known, this model can be solved via the Bethe Ansatz. For a finite
number of particles $N$ in a periodic system of length $L$, eigenstates are fully characterized by a set of real
rapidities $\lambda_j$ satisfying the Bethe equations
\begin{equation}
e^{i\lambda_j L} = \prod_{l \neq j} \frac{\lambda_j - \lambda_l + ic}{\lambda_j - \lambda_l - ic}, \hspace{5mm} j = 1, ..., N.
\label{eq:BAELiebLin}
\end{equation}
The eigenvalues of $\hat{N}$, $\hat{P}$ and $H_0$ on an eigenstate $| \{ \lambda_j \}\rangle$ are respectively $Q_0 = N = \sum_j \lambda_j^0$, $Q_1 = \sum_j \lambda_j$ and $Q_2 = \sum_j \lambda_j^2$. The logic extends to all higher conserved charges $\hat{Q}_n$, whose eigenvalues are simply given by the power sum symmetric polynomials
\begin{equation}
\hat{Q}_n | \{ \lambda \} \rangle = Q_n |\{ \lambda \} \rangle, \hspace{10mm} Q_n \equiv \sum_j \lambda_j^n.
\label{eq:Qneval}
\end{equation} 
The generalized Hamiltonian is therefore diagonalized according to
\begin{equation}
H (\{ \beta \}) | \{ \lambda \} \rangle = E(\{ \beta \}) | \{ \lambda \} \rangle, \hspace{10mm}
E(\{ \beta \}| \{ \lambda \}) = \sum_{n=0}^\infty \sum_{j=1}^N \beta_n \lambda_j^n \equiv \sum_{j=1}^N \varepsilon_0(\lambda_j)
\label{eq:Egen}
\end{equation}
in which we have defined the function 
\begin{equation}
\varepsilon_0 (\lambda) \equiv \sum_{n=0}^\infty \beta_n \lambda^n
\label{eq:vareps0}
\end{equation}
by interpreting the coefficients $\beta_n$ as those of its power series.

For a generic eigenstate in the thermodynamic limit, one can approximate the distribution of rapidities by a 
continuous density function $\rho(\lambda)$ taking positive (or zero) values for all $\lambda$, 
the Bethe equations (\ref{eq:BAELiebLin}) transforming into 
an integral equation \cite{LiebLiniger_1963_PR}:
\begin{equation}\label{eq:BE}
\rho(\lambda) + \rho_h(\lambda) = \frac{1}{2\pi} + a_2 * \rho(\lambda), \hspace{20mm}
a_2(\lambda) \equiv \frac{1}{\pi} \frac{c}{\lambda^2 +c^2}
\end{equation}
in which we use the standard convolution notation $f * g (\lambda) \equiv \int_{-\infty}^\infty d\lambda' f(\lambda - \lambda') g (\lambda')$. The function $\rho_h (\lambda)$ then represents the density of available quantum numbers (holes).
For a given $\rho(\lambda)$ one can then easily compute the expectation values of the conserved charges as \cite{Korepin1993}
\begin{equation}
Q_n  = L \int_{-\infty}^{\infty} d\lambda \lambda^n \rho(\lambda).
%\quad   Q_n \equiv \langle \hat{Q}_n \rangle  \quad n=0,1,2,\ldots
\end{equation}

Following the arguments given by Yang and Yang \cite{YangYang1969}, we now consider a generalized partition
function
\begin{equation}\label{eq:gPartition}
 \mathcal{Z} =\int \mathcal{D}[\rho] e^{-G[\rho,\rho_h[\rho]]}
\end{equation}
(note that we have explicitly written $\rho_h$ as a functional of $\rho$ by using the Bethe equations (\ref{eq:BAELiebLin}))
in which the measure is given by a generalized Gibbs `free energy' 
(although this is strictly speaking not an energy anymore, but a dimensionless quantity) functional
\begin{equation}
\label{eq:G}
G[\rho,\rho_h] =  \sum_{n=0} \beta_n Q_n - S[\rho, \rho_h]
\end{equation}
% &= L \int_{-\infty}^{\infty} \rho(k) \left[\sum_{j} \mu_j k^j - \ln(\rho(k)+\rho_h(k)) + \ln \rho(k)  \right] + \rho_h(k) \left[ - \ln(\rho(k)+\rho_h(k)) + \ln \rho_h(k)  \right] dk
%\end{align}
where the entropy is given to leading order in system size as \cite{YangYang1969,Dorlas1989}
%\footnote{%As  Yang and Yang  themselves point out, this derivation (unlike the rest of their paper), is far from rigorous. 
%The derivation of Yang and Yang is somewhat non-rigorous; a more complete proof was given in \cite{Dorlas1989}.}
%It is, in fact, an ingenious elaboration of the derivation given by Landau and Lifshitz for the nonequilibrium
%entropy density of a free quantum gas.
\begin{equation}\label{eq:entropy}
S = L \int_{-\infty}^{\infty} d\lambda \left[ (\rho+\rho_h) \ln(\rho+\rho_h) - \rho \ln \rho - \rho_h \ln \rho_h \right].
\end{equation}
In the thermodynamic limit we can evaluate the partition function in the saddle-point approximation. 
This leads to the saddle-point condition
\begin{equation} \label{SPC_GGE}
\ln \frac{\rho_h(\lambda)}{\rho(\lambda)} = \sum_{n} \beta_n \lambda^n - a_2 * \ln [1+ \rho(\lambda)/\rho_h(\lambda)].
\end{equation}
By proceeding as usual and defining the function
\begin{equation}
\varepsilon(\lambda) = \ln \frac{\rho_h(\lambda)}{\rho(\lambda)},% \qquad \varepsilon_0(k)=  \sum_{j} \mu_j k^j 
\end{equation}
the saddle-point condition can be rewritten as
\begin{equation}\label{eq:gTBA}
\varepsilon(\lambda) + a_2 * \ln (1+ e^{-\varepsilon(\lambda)}) = \varepsilon_0(\lambda)
\end{equation}
which we call the generalized TBA equation. This is a straightforward generalization of the usual TBA equations,
the only difference being that the driving function $\varepsilon_0(\lambda)$ is given by the generic polynomial
(\ref{eq:vareps0}). One could even think of lifting the restriction of $\varepsilon_0$ to polynomial functions,
and consider functions with isolated singularities. In view of the applications we have in mind, we will 
however not consider these more general cases here.

The equilibrium ({\it i.e.} saddle-point) state of the generalized distribution (\ref{eq:G})
is completely determined by \eqref{eq:BE} and \eqref{eq:gTBA}. 
One can rewrite \eqref{eq:BE} to eliminate $\rho_h(\lambda)$ obtaining
\begin{equation}\label{eq:gTBA2}
\rho(\lambda) = \vartheta(\lambda) \left( \frac{1}{2 \pi} + a_2 * \rho(\lambda) \right)
\end{equation}
in which $\vartheta(\lambda)$ is as usual called the Fermi weight and is defined as
\begin{equation}
\vartheta(\lambda) =  \frac{\rho(\lambda)}{\rho(\lambda)+\rho_h(\lambda)} =  \frac{1}{1+ e^{\varepsilon(\lambda)}}.
\end{equation}
As in the usual case, many expectation values only depend on $\rho(\lambda)$ and $\vartheta(\lambda)$ (see for instance  \cite{Kormos_PRL_2009,Kormos_PRA_2010,Pozsgay2010,Kormos2011,Pozsgay2011}); 
these results can therefore automatically be generalized to solutions of the generalized TBA equations.

For a given set  of $\beta_n$ the saddle-point state is completely determined by 
equations \eqref{eq:gTBA} and  \eqref{eq:gTBA2}. 
Conversely, for a given $\varepsilon(\lambda)$ the Lagrange multipliers can be determined explicitly from \eqref{eq:gTBA},
\begin{equation}\label{eq:Lagrangemultipliers}
\beta_n  = \left. \frac{\partial^n}{\partial \lambda^n} \left( \varepsilon(\lambda)+ a_2 * \ln [1+ e^{-\varepsilon(\lambda)}] \right) \right|_{\lambda=0}.
\end{equation}

One can proceed in either of two ways, depending on the available data. 
Either for a given set $\beta_n$ one solves for $\varepsilon(\lambda)$ using \eqref{eq:gTBA} and from there 
finds $\rho(\lambda)$ using \eqref{eq:gTBA2}. This would be the procedure to follow for example in the case where
the explicit values of the $\beta_n$ correspond to specific `user-defined' perturbations to the original Hamiltonian
\cite{Tsvelik_1990,Frahm_1992,Trippe_2007}, or to situations in which (by some miracle) the generalized inverse
temperatures of the GGE ensemble are known. 

A dual interpretation and use of the generalized TBA equations consists, in fact, in starting from a given 
$\rho(\lambda)$, and solving for $\varepsilon(\lambda)$.  
The Lagrange multipliers, which ultimately encode all characteristics of the saddle-point state, 
can then be computed via \eqref{eq:Lagrangemultipliers}.
While it might seem strange to expect the distribution $\rho(\lambda)$ to be given as input, 
this case is in fact the one which occurs when one uses (numerical) renormalization around a 
Bethe Ansatz-solvable point \cite{2012_Caux_Konik_TBP}. This case in fact represents possibly the most immediately
useful application of the gTBA equations, since the knowledge of $\rho(\lambda)$ then 
allows to explicitly compute GGE predictions {\it without} explicit knowledge of the set $\{\beta_j\}$, by
using the generalized TBA equations to relate $\rho(\lambda)$ to the physical distribution $\varepsilon(\lambda)$ actually
used in the averaging. In the case of the Lieb-Liniger model, in fact, the problems with attempting a direct
implementation of the GGE according to the prescription discussed in the introduction, are compounded by the fact 
that the conserved charges are not properly normal-ordered objects \cite{Davies1990,Davies1990b} whose expectation
values can easily be computed using standard methods. The alternate route which is offered by the generalized TBA equations
is thus actually the only practical one available. We refer the reader to \cite{2012_Caux_Konik_TBP} for a parallel
exposition of the gTBA approach together with its explicit implementation in specific quench cases.

\section{Solving the generalized TBA equations}\label{eq:gTBA_solution}
We can show that a solution of the generalized TBA equation \eqref{eq:gTBA} exists 
and can be found via iteration, in complete parallel to the traditional case, provided that two very simple
conditions are fulfilled. The bare energy 
\begin{equation}
\varepsilon_0(\lambda) = \sum_{n} \beta_n \lambda^n
\end{equation}
should be a) bounded from below, and b) be such that $\lim_{\lambda \rightarrow  \pm \infty} \varepsilon_0(\lambda) = +\infty$.

The proof we give is essentially a repetition of the traditional one offered in \cite{YangYang1969}.
Let us construct the following sequence of functions 
\begin{equation}\label{eq:iterate}
\varepsilon_{n+1}(\lambda) = \varepsilon_0(\lambda)  + A[\varepsilon_n(\lambda)] 
\end{equation}
where
\begin{equation}\label{eq:A_func}
A[\varepsilon(\lambda)]  = - \int_{-\infty}^{\infty} d\lambda' a_2 (\lambda-\lambda') \ln \left( 1 + e^{-\varepsilon(\lambda')} \right).
\end{equation}
The proof contains two steps. First we will show that for every $\lambda$, the sequence of functions is strictly decreasing,
\begin{equation}\label{eq:sequence}
\varepsilon_0(\lambda) > \varepsilon_1(\lambda) > \ldots > \varepsilon_n(\lambda) > \varepsilon_{n+1}(\lambda) > \ldots
\end{equation}
Secondly, we will show that this sequence is bounded from below so that the limit
\begin{equation}
\varepsilon(\lambda)  = \lim_{n\rightarrow \infty} \varepsilon_n(\lambda) \quad 
\end{equation}
exists and is a solution of \eqref{eq:gTBA}. 

From \eqref{eq:A_func} we see that $A[\varepsilon(\lambda)] <0 $ for all $\varepsilon(\lambda)$. 
In order to show that $A[\varepsilon_{n+1}(\lambda)] < A[\varepsilon_n(\lambda)]$ we consider
\begin{equation}\label{eq:dA_func}
\delta A[\varepsilon_n(\lambda)] = \int_{-\infty}^{\infty} d\lambda' a_2(\lambda-\lambda') 
\frac{1}{1+e^{\varepsilon_n(\lambda')}} \delta \varepsilon_n(\lambda').
\end{equation}
For $\delta \varepsilon_n(\lambda) <0$ we have $\delta A[\varepsilon_n(\lambda)]<0$ hence we arrive at \eqref{eq:sequence}.
To prove that $\varepsilon(\lambda)$ is bounded from below is more complicated. First we specialize to cases
in which $\varepsilon_0(\lambda)$ is symmetric in $\lambda$ and is monotonically increasing for positive $\lambda$. 
By considering
\begin{equation}
\frac{d \varepsilon_{n+1}(\lambda)}{d\lambda} = \frac{d \varepsilon_0(\lambda)}{d\lambda} 
+ \int_{-\infty}^{\infty}d\lambda' a_2(\lambda-\lambda') \frac{1}{1+e^{\varepsilon_n(\lambda')}} \frac{d\varepsilon_n(\lambda')}{d\lambda'} 
\end{equation}
we can prove by induction that $\varepsilon_n(\lambda)$ is also symmetric in $\lambda$ and monotonically 
increasing for positive $\lambda$. From this follows that  $A[\varepsilon_n(\lambda)]$ is monotonically 
increasing as a function of $\lambda$, and goes to zero in the limit $\lambda \rightarrow \infty$. 
Using this fact we can write the following inequality
\begin{equation}\label{eq:inequality}
\varepsilon_n(\lambda) = \varepsilon_0(\lambda) + A[\varepsilon_{n-1}(\lambda)]
  \geq \varepsilon_0(\lambda) - \varepsilon_0(0) + \varepsilon_n(0)
\end{equation}
where the equality holds for $\lambda=0$. Using this inequality we can write another inequality using \eqref{eq:iterate} and \eqref{eq:dA_func}
\begin{equation}
\varepsilon_{n+1}(0)  \geq  \varepsilon_0(0) - \int_{-\infty}^{\infty} d\lambda' a_2(0-\lambda') \ln(1+e^{-(\varepsilon_0(\lambda') - \varepsilon_0(0) + \varepsilon_n(0))}).
\end{equation}
Next, we define the function
\begin{align}
f(x) &= \varepsilon_0(0) - \int_{-\infty}^{\infty} d\lambda' a_2(0-\lambda') \ln(1+e^{-(\varepsilon_0(\lambda')-\varepsilon_0(0) +x)}) \\
 &= \varepsilon_0(0) +x - \int_{-\infty}^{\infty} d\lambda' a_2(0-\lambda') \ln(e^{x}+e^{-(\varepsilon_0(\lambda')-\varepsilon_0(0))})
\end{align}
so that \eqref{eq:inequality} can be written as $\varepsilon_{n+1}(0) \geq f(\varepsilon_n(0))$. 
The function $f(x)$ increases monotonically and is bounded from above, $f(x) \leq \varepsilon_0(0)$. 
One can also show that $f(x)-x$ decreases monotonically and that its image is $(-\infty,\infty)$. 
Thus the equation $f(x_0) -x_0 =0$ has a unique solution. For a given $\varepsilon_0(\lambda)$ we can now 
determine $x_0$ from the equation
\begin{equation}
\varepsilon_0(0) =   \int_{-\infty}^{\infty} d\lambda' a_2(0-\lambda') \ln(e^{x_0}+e^{-(\varepsilon_0(\lambda')-\varepsilon(0))}).
\end{equation}
We shall now prove by induction that $\varepsilon_n(0) \geq x_0$ for all $n$. 
First one notes that $\varepsilon_0(0) \geq f(x_0) = x_0$. Next, suppose that $\varepsilon_n(0) \geq x_0$.
One then has from $\varepsilon_{n+1}(0) \geq f(\varepsilon_n(0))$ and the monotonicity of $f(x)$  that
\begin{equation}
\varepsilon_{n+1}(0) \geq f(\varepsilon_n(0)) \geq f(x_0) = x_0.
\end{equation}
This proves the inequality
\begin{equation}
\varepsilon_n(\lambda) \geq \varepsilon_0(\lambda) - \varepsilon_0(0) + x_0 \quad \forall \; \lambda.
\end{equation}
Combining this with \eqref{eq:sequence} we have proved that the solution of \eqref{eq:gTBA} can be found by 
iteration for a symmetric $\varepsilon_0(\lambda)$ which increases monotonically for positive $\lambda$. 
Consider now a general  $\varepsilon_0(\lambda)$ not satisfying these conditions.
Since there obviously exists an $\tilde{\varepsilon}_0(\lambda)$ that is symmetric in $\lambda$, increases monotonically 
for positive $\lambda$ and $\tilde{\varepsilon}_0(\lambda) \leq \varepsilon_0(\lambda)$ for all $\lambda$,
then from \eqref{eq:dA_func} follows that $A[\tilde{\varepsilon}_n(\lambda)] < A[\varepsilon_n(\lambda)] $ 
for all $n$ and $\lambda$. Hence, $\varepsilon(\lambda)$ is bounded from below as well.

\section{Uniqueness of the solution}\label{eq:gTBA_unique}
We can also show that a solution of \eqref{eq:gTBA} %maximizes the partition function 
extremizes the generalized free energy, following the lines of \cite{YangYang1969}. 
We will also show that this solution is unique. 

Consider two solutions of the Bethe equations $\rho_1$ and $\rho_2$. It is clear that $x \rho_1 + (1-x) \rho_2$ for $0 \leq x \leq 1$ is also a solution. Using this property we can define  an action $X(L,\{\beta_n\},\rho)$ by
\begin{equation}
X = L \int_{-\infty}^{\infty} d\lambda \left[ \rho(\lambda) \sum_{n} \beta_n \lambda^n + \rho \ln \rho + \rho_h \ln \rho_h - (\rho+\rho_h) \ln(\rho+\rho_h)\right] 
\end{equation}
and vary it with respect to $\rho$. Consider $\rho=\rho_0+x\rho_1$ where $\rho_0$ and $\rho_1$ are two independent solutions satisfying \eqref{eq:BE}. The variable $x$ takes real values. We can differentiate the action $X$ with respect to $x$
\begin{equation}
\frac{dX}{dx} = L \int_{-\infty}^{\infty} d\lambda \rho_1 \left[  \sum_{n} \beta_n \lambda^n - \varepsilon(\lambda) 
- \int_{-\infty}^{\infty} d\lambda' a_2(\lambda -\lambda') \ln(1+e^{-\varepsilon(\lambda')}) \right].
\end{equation}
Next, using \eqref{eq:gTBA2}, we can compute
\begin{equation}\label{eq:depsilondx}
\frac{\partial \varepsilon}{\partial x} =  \frac{1+e^{-\varepsilon}}{\rho} \left( \frac{1}{1+ e^{\varepsilon}} 
\int_{-\infty}^{\infty} d\lambda' a_2(\lambda-\lambda') \rho_1(\lambda')  - \rho_1\right).
\end{equation}
Now,
\begin{equation}\label{eq:dX2}
\frac{d^2 X}{dx^2} = L \int_{-\infty}^{\infty} d\lambda \rho_1(\lambda) 
\left( -\frac{\partial \varepsilon(\lambda)}{\partial x} + \int_{-\infty}^{\infty} d\lambda' 
\frac{a_2(\lambda-\lambda')}{1+e^{-\varepsilon(\lambda)}} \frac{\partial \varepsilon(\lambda)}{\partial x} \right).
\end{equation}
By first performing the integral over $\lambda$ and then using \eqref{eq:depsilondx} this can be simplified as
\begin{equation}\label{eq:convex_TBA}
\frac{d^2 X}{dx^2}  = L \int_{-\infty}^{\infty} d\lambda \left( \frac{\partial \varepsilon(\lambda) }{\partial x} \right)^2 
\frac{\rho(\lambda)}{1+e^{-\varepsilon(\lambda)}}  >0.
\end{equation}
Since this is true for any $\rho$ we conclude that $X$ is convex, hence it has a unique minimum.

\section{Thermodynamics}
Now that we have established saddle-point (equilibrium) state we can write down the standard thermodynamic relations. First we write generalized free energy \eqref{eq:G} as
\begin{equation}
G =  L \int_{-\infty}^{\infty} \rho(\lambda)(\varepsilon_0(\lambda) -\varepsilon(\lambda) ) -  (\rho(\lambda)+\rho_h(\lambda)) \ln(1+e^{-\varepsilon(\lambda)}) d\lambda.
\end{equation}
At the saddle-point state we use \eqref{eq:gTBA} and \eqref{eq:gTBA2} to further simplify the expression
\begin{align}
G 
%&=  L \int_{-\infty}^{\infty} \rho(k)(a_2 * \ln(1+e^{-\varepsilon(k)}) ) - (a_2 * \rho(k)) \ln(1+e^{-\varepsilon(k)})-\frac{1}{2\pi}  \ln(1+e^{-\varepsilon(k)}) dk\\ \label{eq:G2}
&=  -\frac{1}{2\pi}  L \int_{-\infty}^{\infty}  \ln(1+e^{-\varepsilon(\lambda)}) d\lambda.
\end{align}
%\subsection{Thermodynamic identities}
We can easily show that $G$ satisfies the following thermodynamic identities corresponding to the Hellmann-Feynman theorem
\begin{equation}
\frac{\partial G}{\partial \beta_n} = \langle \hat{Q}_n \rangle.
\end{equation}
We first differentiate \eqref{eq:gTBA} with respect to $\beta_n$
\begin{equation}
\frac{\partial \varepsilon(\lambda)}{\partial \beta_n}= \lambda^n + \int_{-\infty}^{\infty} a_2(\lambda-\mu) \vartheta(\mu) \frac{\partial \varepsilon(\mu)}{\partial \beta_n} d\mu.
\end{equation}
Multiplying both sides with $\rho(\lambda)$ and integrating over $\lambda$ gives
\begin{align}
\int_{-\infty}^{\infty} \frac{\partial \varepsilon(\lambda)}{\partial \beta_n} \rho(\lambda) d\lambda 
%&=\int_{-\infty}^{\infty}  k^n \rho(k) dk + \int_{-\infty}^{\infty} \int_{-\infty}^{\infty} a_2(k-q)  \rho(k) \vartheta(q) \frac{\partial \varepsilon(q)}{\partial \mu_n} dq dk\\
&= \int_{-\infty}^{\infty}  \lambda^n \rho(\lambda) d\lambda + \int_{-\infty}^{\infty}  \left( \rho(\mu)/\vartheta(\mu) - \frac{1}{2\pi} \right) \vartheta(\mu) \frac{\partial \varepsilon(\mu)}{\partial \beta_n} d\mu .
\end{align}
We have used  \eqref{eq:gTBA2} in the second term on the right hand side to eliminate the integral over $\lambda$.
From this follows the identity
\begin{equation}
\frac{\partial G}{\partial \beta_n} = \frac{L}{2\pi}\int_{-\infty}^{\infty} \frac{1}{1+e^{\varepsilon(\lambda)}} \frac{\partial \varepsilon(\lambda)}{\partial \beta_n} d\lambda =\langle \hat{Q}_n \rangle.
\end{equation}
%In case of $\mu_0$ we can obtain a stronger relation
%\begin{equation}
%\frac{\partial \varepsilon(k)}{\partial \mu_0}= 1 + \int_{-\infty}^{\infty} a_2(k-q) \vartheta(q) \frac{\partial \varepsilon(q)}{\partial \mu_0} dq
%\end{equation}
%Since the solution of \eqref{eq:gTBA2} is unique it follows that
%\begin{equation}
%\frac{\partial \varepsilon(k)}{\partial \mu_0} = 2\pi \rho(k) \vartheta(k)^{-1}
%\end{equation}
%\subsection{Susceptibilities}
For the expectation values $\langle \hat{Q}_n \rangle$ we can also define corresponding generalized susceptibilities as
\begin{equation}
\frac{\partial^2 G}{\partial \beta_n^2} =-\left( \langle \hat{Q}_n^2 \rangle - \langle \hat{Q}_n\rangle^2 \right).
\end{equation}
%Applying this to \eqref{eq:G2} we obtain
%\begin{equation}
%\frac{\partial^2 G}{\partial \mu_n^2} = - \frac{L}{2\pi}\int_{-\infty}^{\infty} \frac{e^{\varepsilon(k)}}{(1+e^{\varepsilon(k)})^2} \left(\frac{\partial \varepsilon(k)}{\partial \mu_n}\right)^2 dk   +\frac{L}{2\pi}\int_{-\infty}^{\infty} \frac{1}{1+e^{\varepsilon(k)}} \frac{\partial^2 \varepsilon(k)}{\partial \mu_n^2} dk 
% \end{equation}
%Now by differentiating \eqref{eq:gTBA} twice we obtain
%\begin{align}
%\frac{\partial^2 \varepsilon(k)}{\partial \mu_n^2} =- \int_{-\infty}^{\infty} a_2(k-q) \frac{e^{\varepsilon(q)}}{(1+e^{\varepsilon(q)})^2} \left( \frac{\partial \varepsilon(q)}{\partial \mu_n} \right)^2 dq + \int_{-\infty}^{\infty} a_2(k-q) \frac{1}{1+e^{\varepsilon(q)}} \frac{\partial^2 \varepsilon(q)}{\partial \mu_n^2} dq
%\end{align}
%We multiply with $\rho(k)$ and integrate over $k$  and using similar tricks as we used earlier  we obtain 
%\begin{align}
%\frac{1}{2\pi} \int_{\infty}^{\infty} \frac{\partial^2 \varepsilon(k)}{\partial \mu_n^2}  \frac{1}{1+e^{\varepsilon(k)}} dk &= -  \int_{-\infty}^{\infty} \rho(k) \int_{-\infty}^{\infty} a_2(k-q) \frac{e^{\varepsilon(q)}}{(1+e^{\varepsilon(q)})^2} \left( \frac{\partial \varepsilon(q)}{\partial \mu_n} \right)^2 dq dk\\
%&= -   \int_{-\infty}^{\infty} \left(\rho(q)/\vartheta(q) - \frac{1}{2\pi} \right) \frac{e^{\varepsilon(q)}}{(1+e^{\varepsilon(q)})^2} \left( \frac{\partial \varepsilon(q)}{\partial \mu_n} \right)^2 dq 
%\end{align}
%Combining the results yields
Using similar tricks as before we obtain 
\begin{equation}
\frac{\partial^2 G}{\partial \beta_n^2} =   -L \int_{-\infty}^{\infty} \frac{\rho(\lambda) e^{\varepsilon(\lambda)}}{1+e^{\varepsilon(\lambda)}} \left( \frac{\partial \varepsilon(\lambda)}{\partial \beta_n}  \right)^2 d\lambda < 0,
 \end{equation}
which is essentially identical to (\ref{eq:convex_TBA}). In summary, we can simply state that 
all the usual equilibrium thermodynamic equations remain unchanged
as compared to the usual case, provided one uses the generalized $\varepsilon(\lambda)$ and $\rho(\lambda)$ functions.
%\textbf{Note: this equation is essentially the same as \eqref{eq:convex_TBA}, although the reasoning is slightly different. Maybe we can absorb to section about uniqueness into this one.}

\section{Excitations}
In order to study excitations upon the saddle-point state it is instructive to first go back to a state with a finite number $N$ of particles. To a state with quantum numbers $I_j$ corresponds a set of rapidities $\lambda_j$ that satisfy the Bethe equations
\begin{equation}
\lambda_j L  = 2 \pi I_j - \sum_{i=1}^N \theta(\lambda_j - \lambda_i).
\end{equation}
Now we consider an 'excited' state with quantum numbers $I_j'$ and the rapidities $\lambda_j'$ satisfying different Bethe equations
\begin{equation}
\lambda_j' L = 2\pi I_j' - \sum_{i=1}^N \theta(\lambda_j'-\lambda_i')
\end{equation}
such that $I_j = I_j' \quad \text{except when } j= n$. We make the usual assumption that for all $j \neq n$ the difference between $\lambda_j$ and $\lambda_j'$ is small. One can introduce a shift function $\chi(\lambda)$
\begin{equation}
(\lambda_j'-\lambda_j)L = \chi(\lambda_j) \quad \text{for } j \neq n.
\end{equation}
Going back to the thermodynamic limit one can show using the arguments of Lieb \cite{Lieb_1963} and Yang and Yang \cite{YangYang1969} that $\chi(\lambda)$ is determined by the following integral equation
\begin{equation}
\chi(\lambda) =2\pi \int_{-\infty}^{\infty} a_2(\lambda-\mu) (\chi(\lambda)-\chi(\mu)) \rho(\mu) d\mu + \theta(\lambda-\lambda_n) - \theta(\lambda-\lambda_n').
\end{equation}
Writing the back-flow as $g(\lambda) = \chi(\lambda) (\rho(\lambda)+\rho_h(\lambda))$ we obtain
\begin{equation}\label{eq:Backflow}
 g(\lambda) = \int_{-\infty}^{\infty} a_2(\lambda-\mu) g(\mu)\vartheta(\mu) d\mu + \frac{1}{2\pi}\left(  \theta(\lambda-\lambda_n)-\theta(\lambda-\lambda_n')\right).
\end{equation}
%This is a Fredholm integral equation of the second kind.
The momentum difference and energy (as measured from the basic Lieb-Liniger Hamiltonian) difference between the two states are
\begin{align}\label{eq:DeltaK}
\Delta P &= \sum_j (\lambda_j'-\lambda_j) = \lambda_n' - \lambda_n + \int_{-\infty}^{\infty} g(\lambda) \vartheta(\lambda) d\lambda,\\
\Delta E &= \sum_j (\lambda_j'^2-\lambda_j^2) = \lambda_n'^2 -\lambda_n^2 + \int_{-\infty}^{\infty} 2\lambda  g(\lambda) \vartheta(\lambda) d\lambda.
\end{align}
In the case in which the density matrix Hamiltonian does not coincide with the Schr\"odinger Hamiltonian, we have that
$\Delta E \neq \varepsilon(\lambda_n') - \varepsilon(\lambda_n)$, therefore $\varepsilon(\lambda)$ cannot be interpreted 
as the energy of fundamental excitations anymore, as opposed to the usual case. However, if the Schr\"odinger Hamiltonian
is by definition the generalized Hamiltonian $H = H_{eff} = \sum_{n} \beta_n \hat{Q}_n$, its excitations are
\begin{align}\label{eq:DeltaHeff}
\Delta E_{eff} &= \sum_j (\varepsilon_0(\lambda_j')-\varepsilon_0(\lambda_j)) = \varepsilon_0(\lambda_n') -\varepsilon_0(\lambda_n) + \int_{-\infty}^{\infty} \frac{d\varepsilon_0(\lambda)}{d\lambda} g(\lambda) \vartheta(\lambda) d\lambda
\\ &=  \varepsilon(\lambda_n') -\varepsilon(\lambda_n). 
\end{align}
Hence $\varepsilon(\lambda)$ are the excitations of the effective Hamiltonian.
As in the usual case, it is straightforward to prove that a finite number of simultaneous excitations  
is simply the sum of the individual elementary excitations
\begin{align}
\Delta P(\{\lambda_{p_j}\},\{\lambda_{h_j}\}) &= \sum_j \Delta P(\lambda_{p_j},\lambda_{h_j})\\
\Delta E(\{\lambda_{p_j}\},\{\lambda_{h_j}\}) &= \sum_j \Delta E(\lambda_{p_j},\lambda_{h_j}),
\end{align}
this relation remaining true provided the density of excitations thus created remains zero in the thermodynamic limit.
The whole bulk of knowledge about correlations of the equilibrium Lieb-Liniger gas \cite{Korepin1993} can thus be easily adapted
to the generalized cases.

\section{Conclusion}
Although we have concentrated on the Lieb-Liniger model, similar reasonings are (almost trivially) 
transportable to other models. The generalization of generic TBA equations (with strings, nested, etc.) is straightforward since only
the basic driving function (bare energy) needs to be modified. In cases other than the repulsive Lieb-Liniger model however, 
proving existence and uniqueness becomes even more intractable than in the traditional case, 
although explicit solutions can easily be found in practice.

In summary, we have provided a simple extension of the traditional TBA equations to cases in which higher
conserved charges come into play, either via extended Hamiltonians or via density matrix averaging differing from
the usual Gibbs procedure. For the Lieb-Liniger, existence and uniqueness is proven provided the driving
function fulfills very simple conditions. The use of the generalized TBA equations provides a practical road to the
actual implementation of generalized Gibbs ensemble averages, or other similar ones, by offering an alternate route
which obviates the need to work with the actual expressions for the conserved charges and their associated inverse temperatures.

\section*{Acknowledgements}
We thank R. Konik, T. Fokkema and S. Eli\"ens for useful discussions. We remind the reader that 
the gTBA approach is also developed and used in \cite{2012_Caux_Konik_TBP}. 

This work was supported by the Foundation for Fundamental Research on Matter (FOM) and from the Netherlands 
Organisation for Scientific Research (NWO). 

\bibliography{gTBA_good}
\bibliographystyle{unsrt}

\end{document}